\documentclass[a4paper]{article}

\usepackage{graphicx}
\usepackage{amsmath}
\usepackage{amsfonts}
\usepackage{amssymb}

\begin{document}

\author{G. Labeyrie, C. A. M\"{u}ller, D. S. Wiersma$^{\ast}$,
Ch. Miniatura, R.Kaiser\\
\textit{INLN UMR 6618 CNRS-UNSA}\\
\textit{\ 1361 route des Lucioles, F-06560 Valbonne} \\
$^{\ast}$\textit{European laboratory for non-linear spectroscopy}\\
\textit{and Instituto nazionale per la fisica della materia}\\
\textit{Florence, Italy}}
\title{Observation of coherent backscattering of light by cold atoms}
\maketitle
\begin{abstract}
Coherent backscattering (CBS) of light waves by a random medium is a signature
of interference effects in multiple scattering. This effect has been studied
in many systems ranging from white paint to biological tissues. Recently, we
have observed CBS from a sample of laser-cooled atoms, a scattering medium
with interesting new properties. In this paper we discuss various effects,
which have to be taken into account for a quantitative study of coherent
backscattering of light by cold atoms.
\end{abstract}

\section{Introduction}

A wave propagating in a strongly scattering random medium undergoes many
scattering events and the memory of its initial direction is rapidly lost.
This simple observation applies to many everyday life situations, like driving
a car in thick fog. Understanding the rules of wave propagation in such media
may have some interesting applications e.g. in medical imaging or in
mesoscopic physics.

Since the wave propagation can be seen as a random walk inside the medium, a
diffusion picture seems appropriate. Neglecting all interference phenomena,
one predicts a total transmission of the medium inversely proportional to
sample thickness (Ohm's law). However, interferences may have dramatic
consequences, such as a vanishing diffusion constant : in this situation, the
medium behaves like an insulator (strong or Anderson
localization)\cite{anderson} and its total transmission decreases
exponentially with the sample's thickness. This prediction has triggered a
renewal of interest for the study of multiple scattering, leading to
experiments on strong localization of microwaves\cite{micro} and
light\cite{wiersmaNat}. A more accessible experimental situation is the
so-called weak localization regime, where interferences already hamper the
diffusion process. Coherent backscattering (CBS) is a spectacular
manifestation of interference effects in this multiple scattering regime,
yielding an enhanced scattered intensity around the direction of
backscattering. This phenomenon\ has been observed in a variety of systems
\cite{cbsexp}.

Recently, we observed coherent backscattering of light from a sample of
laser-cooled atoms\cite{labeyrie}. Indeed, multiple scattering of light is
known to exist in such samples since it eventually limits the atomic density
achievable in magneto-optical traps (MOTs) \cite{densMOT}. Direct
manifestations of multiple scattering of light in cold atoms such as
''radiation trapping'' had already been observed \cite{OptTrap}, but our
experiment now allows to probe the \textit{interference} effects in this
situation. In this respect, CBS is a powerful tool to study the properties of
light scattered by cold atoms. Indeed, we observed some striking differences
with what is reported in the literature for classical samples. In order to
understand more precisely the physics underlying these differences, we have to
analyze various effects, such as geometrical or polarization effects, which
could modify the coherent backscattering signal even for classical samples
such as a suspension of TiO2 beads. The goal of this paper is to study such
effects in order to point out behaviors connected to the internal structure of
the atoms.

In section 2, we first recall the basic physics of coherent backscattering,
with a special attention to the parameters that determine the CBS cone shape.
Section 3 is devoted to experiments with classical samples. After describing
the detection setup, we discuss several effects that can affect the signal. We
put an emphasis on the rather non-trivial problem of determining a precise
value of the enhancement factor. Section 4 is dedicated to the experiment with
cold atoms, including a description of the procedure to prepare the sample.

\section{Coherent backscattering}

\subsection{Principle of coherent backscattering}

\begin{figure}
\center
\includegraphics[width=0.75\textwidth]{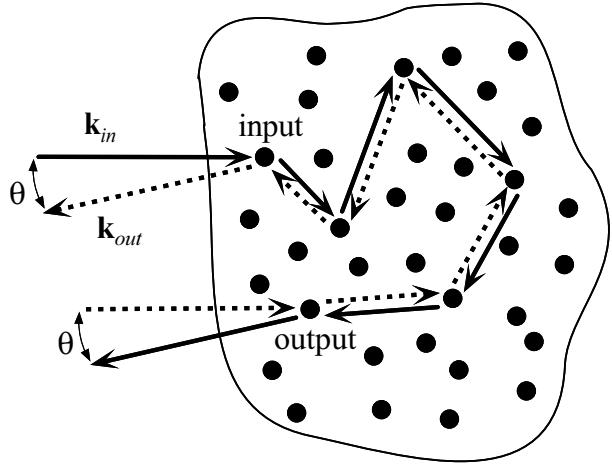}
\caption{The origin of CBS.}
\label{fig1}
\end{figure}

To understand the origin of CBS, let us consider the situation depicted in
fig. \ref{fig1}. 
A sample of randomly distributed scatterers is illuminated by a plane
wave (wavelength in vacuum $\lambda$, wave vector $\mathbf{k}_{\text{in}}$ ).
The quantity of interest is the angular distribution of the scattered light
intensity in the backward direction. We consider here the simplest case of
scalar waves. Some consequences of the vector nature of the light waves will
be discussed in section $3.2$.

\begin{figure}[h]
\center
\includegraphics[width=0.5\textwidth]{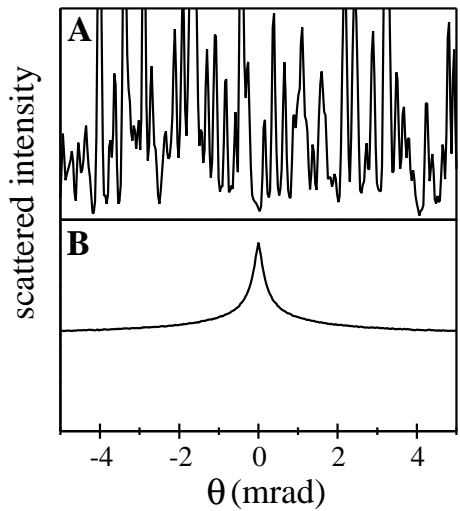}
\caption{Interference effects and configuration average. \textbf{A}
: backscattered intensity for one configuration of the sample (speckle)
\textbf{B} : backscattered intensity after configuration average (CBS cone).}
\label{fig2}
\end{figure}

If the scatterers' respective positions are fixed, the light intensity
scattered at angle $\theta$ results from the interference of many partially
scattered waves and is a fast-varying function of $\theta$. This is the
well-known speckle pattern (see fig~\ref{fig2}\textbf{A}). Speckle is observed whether
the medium is optically thin, with single scattering being dominant, or
optically thick in the multiple scattering regime. Let us now imagine that a
configuration averaging is performed : the respective positions of the
scatterers in the sample are modified and the corresponding different speckle
patterns are summed up, resulting in an averaged intensity distribution. In
experiments, this is obtained either automatically due to the scatterer's
motion (e.g. in liquid samples), or by moving the sample so that different
configurations are probed. As a result of this averaging process, we expect
the speckle pattern to smooth out to give a relatively angle-independent
intensity distribution. The main argument in this explanation is that the
detected field is the coherent sum of scattered electric fields :%

\begin{equation}
\mathbf{E}=\underset{j}{\sum}\mathbf{E}_{\text{j}}\exp\left(  i\varphi
_{\text{j}}\right)
\end{equation}
The average detected intensity will then be :%

\begin{equation}
\left\langle I\right\rangle =\left\langle \left|  \underset{j}{\sum}%
\mathbf{E}_{\text{j}}\exp\left(  i\varphi_{\text{j}}\right)  \right|
^{2}\right\rangle =\underset{j}{\sum}\left\langle \left|  \mathbf{E}%
_{\text{j}}\right|  ^{2}\right\rangle +\underset{j\neq k}{\sum}\left\langle
\mathbf{E}_{\text{j}}\cdot\mathbf{E}_{\text{k}}^{\ast}\exp\left(
i\varphi_{\text{j}}-i\varphi_{\text{k}}\right)  \right\rangle
\label{intensitediff}%
\end{equation}
where the brackets denote configuration averaging. A first approach would be
to suppose that the phases $\varphi_{\text{j}}$ and $\varphi_{\text{k}}$ are
uncorrelated random variables, which yields an interference term equal to zero :%

\begin{equation}
\left\langle I\right\rangle =\left\langle \underset{j}{\sum}\left|
\mathbf{E}_{\text{j}}\right|  ^{2}\right\rangle
\end{equation}
However, this argument is wrong if the interference arises from two
\textit{correlated} fields. Such correlations can be very important in the
case of spatial correlation of the scatterers, as e.g. for Bragg scattering in
crystals.\ But even if there is no correlation in the position of the
scatterers, the fields $\mathbf{E}_{\text{j}}$ and $\mathbf{E}_{\text{k}%
}^{\ast}$ can be correlated. In particular, this is the case for
backscattering in the multiple scattering regime.

Indeed, let us consider for every scattering path (yielding some
backscattering), the \textit{reverse} path as represented on fig~\ref{fig1}. This
reverse path (dotted arrows) involves the same scattering sequence as the
''direct'' path (solid arrows), but in inverse order. The geometrical phase
difference between waves following these two paths is :%

\begin{equation}
\Delta\varphi=\left(  \mathbf{k}_{\text{in}}+\mathbf{k}_{\text{out}}\right)
\cdot\left(  \mathbf{r}_{\text{in}}-\mathbf{r}_{\text{out}}\right)
\label{phase}%
\end{equation}
where $\mathbf{r}_{\text{in}}$ and $\mathbf{r}_{\text{out}}$ are the vector
positions of \ the first and last scatterers involved in the path (denoted by
''input'' and ''output'' in fig~\ref{fig1}). One can thus see that if the relative
position of the scatterers is randomly changing the phase difference is
generally also a random parameter and the corresponding interference terms in
eq.(\ref{intensitediff}) will be cancelled. However, for the particular case
of backscattering ($\mathbf{k}_{\text{in}}+\mathbf{k}_{\text{out}}=0$) this
phase difference is always zero, \textit{regardless of the specific scattering
path} considered. Thus, the two waves following the reverse paths of
fig~\ref{fig1}  
always add up constructively in the backscattering direction, and this
interference survives the averaging process (this property is of course not
verified for $\theta\neq0$, where the interferences vanish). The remaining
terms in eq.(\ref{intensitediff}) arise from interference between
\textit{distinct }paths and are obviously zero since the fields are not
correlated in this case. As a result, the averaged intensity distribution
exhibits a peak centered at $\theta=0$, known as the CBS cone. This is
illustrated on fig~\ref{fig2}, where the intensity distributions for a given fixed
configuration (\textbf{A}) and after configuration average (\textbf{B}) are
recorded.\ Note that, in the case of a single configuration (speckle), one
does not necessarily have\ a constructive interference in the backscattering
direction.

The ratio of the configuration-averaged scattered intensity at $\theta=0$
(exact backscattering) to the ''incoherent background'' obtained at large
angle is known as the CBS enhancement factor. If the amplitudes of the reverse
paths, which interfere in the backscattering direction, are equal, the
enhancement factor equals 2. However, this property is verified only if the
single scattering light, which does not contribute to CBS, is removed from the
detected signal. We will see in section 3.2 how this can be achieved by
selecting the appropriate polarization channel.

\subsection{Cone shape}

The enhanced backscattering described above relies on the constructive
interference between reverse paths. One can make an analogy with a Young's
interference experiment, where two diffracting slits would be positioned in
place of the ''input'' and ''output'' scatterers (see fig~\ref{fig1}). If the slits are
backlit with a plane wave (of wave vector $-\mathbf{k}_{\text{in}}$), the
interference produces a sinusoidal fringe pattern in the far field, with a
maximum intensity at $\theta=0$ and a fringe spacing inversely proportional to
the \textit{transverse} spacing between the scatterers (this is valid only for
small values of $\theta$). The total configuration-averaged intensity
distribution is obtained by summing up incoherently the fringe patterns
corresponding to all the possible scattering paths in the sample. This
incoherent sum accounts for the fact that interferences between waves
following distinct paths do not survive the configuration average. Since the
fringe patterns all have a ''bright'' fringe at $\theta=0$, the total
intensity is maximum there, and decreases to an ''incoherent background''
value within an angular range $\Delta\theta\sim\lambda/d$, where $d$ is the
average transverse distance between slits (this is similar to the zero
path-difference fringe observed in Michelson interferometers with white
light). This analogy thus shows why the average light intensity is increased
around the backscattering direction, and relates the angular width of the peak
to the inverse of the distance between entering and exit points of the light
in the sample.

More precisely, in the case of a semi-infinite medium and for scalar waves,
the FWHM of the coherent backscattering cone is given by \cite{Maynard,
depol}\emph{\ }:%

\begin{equation}
\Delta\theta_{CBS}\approx\frac{0.7}{kl^{\ast}}\label{FWHMcbs}%
\end{equation}
where $k$ is the wave vector in the scattering medium, and $l^{\ast}$ is the
\textit{transport} mean free path. The transport mean free path describes the
distance necessary, on average, for the initial direction of propagation to be
scrambled (which is of course essential to observe backscattering). It is
related to the \textit{scattering} mean free path $l$ (mean distance between
two scattering events) by:%

\begin{equation}
l^{\ast}=\frac{l}{1-\left\langle \cos\theta\right\rangle }%
\end{equation}
where $\theta$ is the angle between the incident and scattered light (for a
single scatterer), and the brackets denote the average over the radiation
pattern of the scatterer. Thus, if $\left\langle \cos\theta\right\rangle =0 $
the scattering and transport mean free paths are identical. Note that this
condition does not imply that the radiation pattern is isotropic (think for
instance of the dipole radiation pattern).

\begin{figure}[h]
\center
\includegraphics[width=\textwidth]{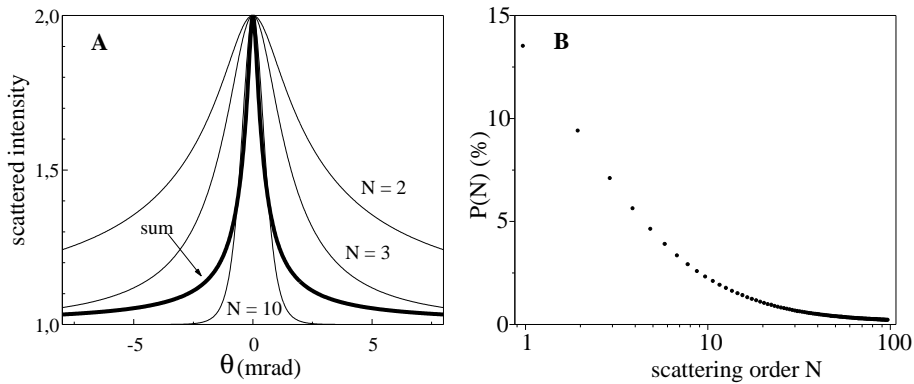}
\caption{Contribution of different scattering orders in a slab of
optical thickness $b=12$. \textbf{A} : CBS cones for $N=2,3,10$ and the sum of
the first 80 orders \textbf{B} : contribution of each order to the
backscattered intensity.}
\label{fig3}
\end{figure}

As it was evidenced with the Young's slits analogy, the width of the coherent
backscattering cone depends on the mean distance between the first and last
scatterers. This distance will of course increase with the scattering order
$N$ (number of scattering events) involved, so higher orders will yield
narrower cones. For \textit{large} scattering orders ($N\gg1$) the propagation
can be described as a random walk of step $l^{\ast} $, and the average
distance between the input and output scatterers grows as $\sqrt{N}%
$\textit{\ }$l^{\ast}$ (diffusion approximation). In a semi-infinite medium
where all scattering orders contribute, the total CBS cone is obtained by
adding up the cones associated to each order. This implies to evaluate the
weight $P(N)$ of each scattering order. Due to the presence of very high
scattering orders (giving very narrow cones), the actual shape of the cone
around the tip is triangular \cite{wiersma}. The resulting angular FWHM is
given by eq.(\ref{FWHMcbs}). The relationship between cone width and
scattering order is illustrated on fig~\ref{fig3}\textbf{A}, where are plotted the CBS
cones associated to $N=2,3,10$ (thin lines) and the sum of all the
contributions up to $N=80$ (bold line), in the case of a slab of non-absorbing
medium of optical thickness $b=12$.

The scattered intensity is plotted as a function of the normalized
backscattering angle $\theta kl^{\ast}$. These curves are obtained with a
rigorous theory \cite{vanDerMark} for scalar waves, which does not rely on the
diffusion approximation. Each cone is scaled by its own incoherent background
for better comparison of the widths, so the respective amplitudes of the
different orders do not appear on this plot. It can be seen that the width of
the cone decreases as the scattering order increases (the double scattering
cone is approximately 10 times broader than the ''total'' peak). The shape is
also clearly affected, for instance in the ''wings'' of the cones ($\theta
kl^{\ast}\gg1$) : for $N=2$, the scattered intensity decreases as $1/\theta$
while\ the sum of all the other contributions decreases as $1/\theta^{2}$
\cite{Bart}. On \textbf{B} are plotted the weights $P(N)$ corresponding to
each scattering order. Asymptotically, the weight of the $N^{th}$ order
decreases as $N^{-3/2}$ \cite{vanDerMark}.

Thus, we emphasize that the CBS cone shape is in general determined not only
by the transport mean-free path, as in the case of a semi-infinite medium
(eq.(\ref{FWHMcbs})), but also by the sample geometry through a truncation of
the scattering orders. A similar effect is obtained in the case of an
absorbing medium, where the contribution of long light paths is reduced.

\section{Experiments with classical scatterers}

We now turn to the description of CBS experiments using classical samples such
as milk, suspensions of TiO$_{2}$ particles, or teflon. We discuss several
effects that can affect the CBS signal.

\subsection{Description of the experimental setup}

\begin{figure}[h]
\center
\includegraphics[width=0.75\textwidth]{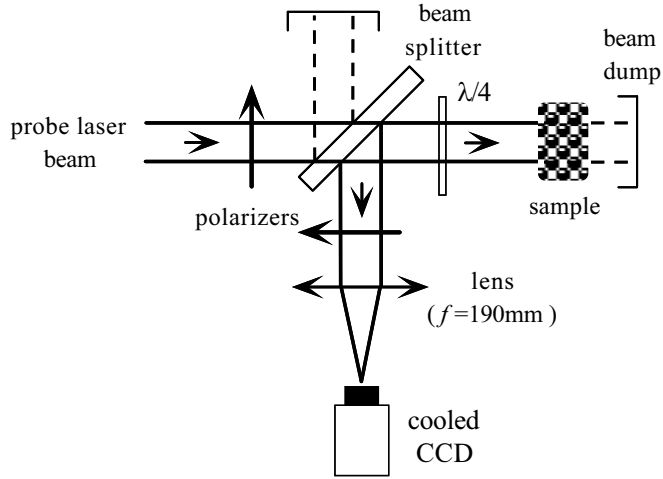}
\caption{CBS detection setup.}
\label{fig4}
\end{figure}

The CBS detection setup used in our experiment is schematically represented on
fig~\ref{fig4}. The sample is illuminated by a collimated laser probe ($1/e^{2}$ waist
7,6 mm). Most of the backscattered light ($\sim$90\%) is reflected by a
beam-splitter, and its angular (far field) distribution is recorded on a
cooled CCD placed in the focal plane of an analysis lens ($f=190$ mm). Since
the focussing is quite critical, the CCD camera is mounted on a translation
stage. By rotating the polarizer and quarter-wave plate, one can select the
polarization channel where the signal is detected (see section $3.1.3$).

As usual in CBS experiments, great care must be taken to shield the detector
against stray light; the alternative paths that can be followed by the light
(incident beam reflected by the beam-splitter and beam transmitted by the
sample) must also be carefully blocked to avoid any unwanted backscattering.
This is achieved by inserting a neutral filter at Brewster angle in the
unwanted beam paths. The residual reflection by the neutral filter is directed
onto a black paper.

Another possible source of stray light originates from reflections inside the
beamsplitter. This is avoided by using a beamsplitter with a small wedge ($5%
{{}^\circ}%
$). This beamsplitter has different reflection coefficients for s- and
p-polarized light (polarization orthogonal or parallel to the plane of
incidence respectively), which have to be accounted for when comparing data in
different polarization channels. Another consideration is the angular response
of the detection optics (quarter-wave plate + beam-splitter + polarizer),
which should be sufficiently flat within the angular field of observation to
avoid deformation of the background level. However, because of our small
detection angular range ($\simeq15$ mrad), this effect is negligible in our case.

The use of a cooled CCD with low thermal (and readout) noise allows for long
integration times\ yielding improved signal-to-noise ratio. This is also
convenient to record CBS cones from self-averaging samples such as milk or a
suspension of TiO$_{2}$ particles, where the CCD camera integrates the
scattered light for several tens of ms up to several minutes, depending on the
time scale of the motion of the scatterers.

\subsubsection{Angular resolution}

To detect CBS cones with a small angular width, it is important to avoid
convolution due to the residual divergence of the incident laser beam and to
aberrations of the detection optics. The probe beam collimation is achieved
using a telescope including a spatial filter, and shear plate interferometry
\cite{shearPlate} as a diagnostic technique. The diffraction limit
corresponding to our beam waist size is $\Delta\theta_{\text{diff}}\simeq0.03$
mrad FWHM, below the resolution limit due to the CCD's pixel size
$\Delta\theta_{\text{pix}}\simeq0.05$ mrad. For technical reasons linked to
the experiment with cold atoms, the actual detection optics is more
complicated than represented on fig~\ref{fig4} and includes an image transport system
between the focal plane of the analysis lens and the CCD (see fig~\ref{fig12}). All the
lenses in the detection system are achromatic doublets to minimize aberrations.

The most direct way to estimate the effective angular resolution of the
experimental setup is to record the CBS cone from a liquid sample, milk for
instance, which is gradually diluted to increase the scattering mean free path
(thus reducing the width of the cone). Once the cone becomes narrower than the
angular transfer function of the apparatus, the observed signal is strongly
reduced due to convolution and its width is essentially that of the transfer
function. Using this procedure, we find an effective angular resolution
$\Delta\theta_{\text{res}}\simeq0.1$ mrad. We believe this value results from
residual aberrations in the optical system. Knowing the effective resolution,
it is then possible to compute the broadening and reduction of the CBS cone
due to convolution.

Although it has many advantages, the choice of a CCD also implies that the
angular dynamics of our detection is somewhat limited (the CCD has
$770\times512$ pixels) compared to, for instance, the system of
ref\cite{facteur2}. Because of the far-reaching wings of typical CBS\ cones,
it is difficult to have at the same time an angular magnification (determined
by the focal length of the analysis lens) good enough to look at the shape of
the cone around the tip, and an angular field wide enough to see the wings.

\subsubsection{Signal acquisition and treatment}

Here we describe our standard procedure to obtain a CBS cone profile such as
that shown on fig~\ref{fig14}. First an image of the CBS cone is recorded. The
configuration average is performed using a small rotor (solid samples) or
simply by the motion of the scatterers (liquids, cold atoms). A typical
integration time is 20 s. Then, a second ''background'' exposure is taken
without sample, and subtracted from the signal to remove residual stray light.
This step will be discussed in more details in the case of an atomic sample.

Once the image is obtained, a cross-section is taken to obtain a profile.
However, in the case of a noisy signal, we perform an angular average on the
CCD image to smoothen the CBS profile : the center of the CBS peak is
pinpointed, and a number of different cross-sections passing through this
center are averaged to give the final signal. We emphasize that this technique
can only be employed if the cone is isotropic, which is the case only in
certain polarization channels (see section $4.3.3$). We checked that, in the
appropriate channels, this procedure yields the same profile as when using a
simple cross-section.

The remaining problem is the determination of the enhancement factor, which
implies an estimation of the level of the incoherent background. As already
mentioned, the wings of the cone are quite wide and a direct measurement of
this background level is difficult. Thus, we fit the experimental profile with
a sum of four lorentzian curves, all centered on $\theta=0$ but with widths
and heights as free parameters. The value of the background is also returned
by the fit and used to determine the enhancement factor. This empirical
approach allows to fit, using the same procedure, different cone shapes whose
analytical expressions are not known. To estimate its accuracy, we applied the
technique to two different theoretical cone shapes : a cone from a
semi-infinite medium (diffusion theory, $I\left(  \theta\right)
\varpropto1/\theta^{2}$ for $\theta\ \gg\lambda/l^{\ast}$) and a
double-scattering cone ($I\left(  \theta\right)  \varpropto1/\theta$ for
$\theta\gg\lambda/l^{\ast}$). For an angular field of detection about 20 times
wider than the FWHM of the cones (typical experimental situation), the error
on the enhancement factor is below 1\%. The actual uncertainty on the
enhancement factor originates from the fact that our smoothing procedure does
not improve the signal-to-noise ratio at the cone tip, because this particular
point is common to all the profiles averaged.\ To reduce the uncertainty, we
average the signal from a few neighboring pixels around the center of the
cone, but the improvement is limited since the corresponding angular range
must remain smaller than the resolution. We finally estimate the uncertainty
on the enhancement factor $f_{e}$ to be around $f_{e}\pm0.01 $.

\subsection{Polarization effects}

An important aspect of all\ coherent backscattering experiments with light is
the vector nature of the scattered wave, i.e. the polarization of the light.
Thus, controlling the incident and detected polarizations is essential in
these experiments.

For a linear incident polarization (quarter-wave plate removed), we
record\ (by rotating the detection polarizer) the scattered light either with
linear polarization parallel (''parallel'' channel \ or $lin$ $//$ $lin$) or
orthogonal (''orthogonal'' channel or $lin\perp lin$) to the incident one. We
also use a circular incident polarization by inserting the quarter-wave plate
between the beam-splitter and the sample. In the ''helicity preserving''
channel (denoted $h$ $//$ $h$) the detected polarization is circular with the
same helicity (sign of rotation of the electric field referenced to the
direction of wave propagation) as the incident one : in this channel, no light
is detected in the case of\ the back-reflection from a mirror. The
''orthogonal helicity'' channel ($h\perp h$) is obtained for a detected
circular polarization orthogonal to the previous one. When defining the
polarization by referring to a \textit{fixed} axis (as one usually does in the
atomic physics community), an incident $\sigma^{+}$\ light would remain
$\sigma^{+}$ by reflection from a mirror. The $h\perp h$ channel is thus a
$\sigma^{+}/\sigma^{+}$\ channel, and the $h$ $//$ $h$ channel corresponds to
a polarization flip from $\sigma^{+}$ to $\sigma^{-}$.

The choice of the appropriate polarization channel makes it possible, at least
for some categories of scatterers, to remove the single scattering
contribution to the detected light. Indeed, single scattering does not
contribute to CBS but adds up to the signal as a background, and thus reduces
the apparent enhancement factor (defined as the ratio of the detected
intensities at $\theta=0$ and $\theta\gg\lambda/l^{\ast}$). In the case of
single (back)scattering, ''spherical'' scatterers e.g. Rayleigh (size
$a<\lambda$) and spherical Mie ($a\gtrsim\lambda$) scatterers behave like
mirrors : they flip the helicity of circularly-polarized light. Thus, CBS
experiments are usually performed in the $h$ $//$ $h$ channel where the single
scattering contribution is rejected. Furthermore, the reciprocity principle
\cite{reciprocity} can be used in this channel, and predicts an enhancement
factor of 2. In the case e.g. of non-spherical scatterers, the single
scattering contribution is present even in the $h$ $// $ $h$ channel and the
expected enhancement factor is smaller than 2 \cite{mishchenko}.

\begin{figure}[h]
\center
\includegraphics[width=\textwidth]{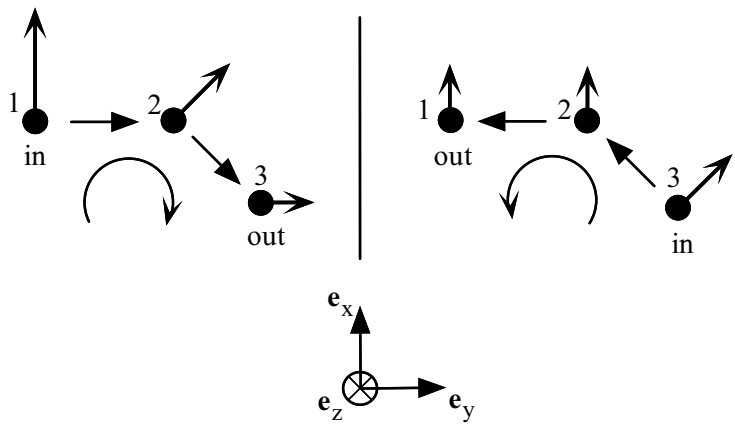}
\caption{Reduction of enhancement factor in the $lin\perp lin$ channel.}
\label{fig5}
\end{figure}

Polarization can also affect the enhancement factor through more subtle ways.
This is illustrated on fig~\ref{fig5} with the example of $N=3$ scattering and dipole
scatterers.

The incident wave vector is orthogonal to the plane of the figure (parallel to
$\mathbf{e}_{\text{z}}$), where all the scattering events are supposed to take
place. We consider the case of detection in the $lin\perp lin$ channel : the
incident wave polarization is parallel to $\mathbf{e}_{\text{x}}$ and the
detected polarization along $\mathbf{e}_{\text{y}}$. The arrows mark the
polarization of the wave after each scattering. This figure illustrates the
fact that the amplitudes of the reverse paths that interfere to give rise to
the cone are different in this channel : for the path on the left (scattering
sequence $1\rightarrow2\rightarrow3$) some light comes out in the polarization
orthogonal to the incident, while for the reverse sequence ($3\rightarrow
2\rightarrow1$) the projection on the detected polarization is zero. Since the
amplitudes of the two waves are imbalanced, the contrast of the interference
will be reduced and hence the CBS\ enhancement factor. This contrast reduction
effect becomes more effective as the order of scattering $N$ increases. Thus,
in the case of \ ''spherical'' scatterers, the enhancement factor in the
''orthogonal'' channels (linear and circular) is 2 for $N=2$ \cite{depol}, and
decreases fast for higher orders. For aspherical scatterers (e.g. antennas),
the enhancement factor in the orthogonal channels is smaller than 2 even for
$N=2$.

\begin{figure}[h]
\center
\includegraphics[width=0.75\textwidth]{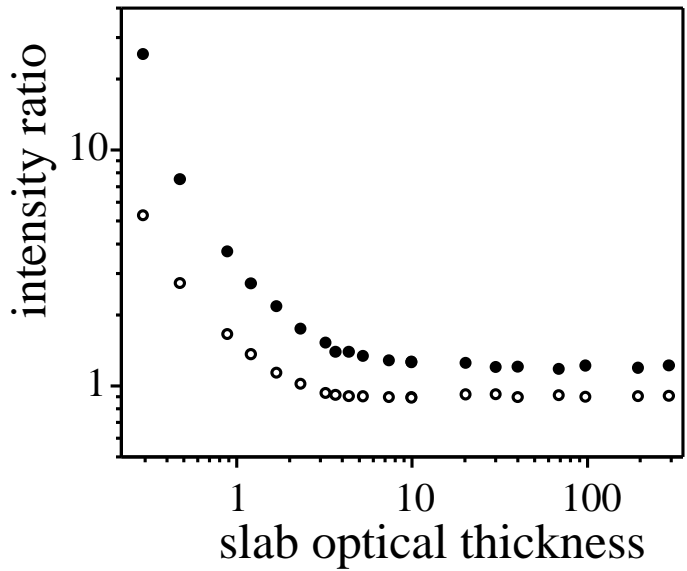}
\caption{Depolarization due to multiple scattering.}
\label{fig6}
\end{figure}

When multiple scattering occurs, the polarization of the incident wave is
rapidly scrambled. This phenomenon is illustrated on fig~\ref{fig6}.

On this plot we reported the ratio of the intensities\ scattered in crossed
channels for an incident circular polarization (ratio $=I_{h//h}/I_{h\perp h}
$, open circles) and linear polarization (ratio $=I_{lin\perp lin}%
/I_{lin//lin}$, full circles), as a function of the optical thickness of the
sample. The sample is a solution of TiO$_{2}$ particles (size $\simeq$ $200$
nm) in a cell with a slab geometry (thickness 7 mm). The concentration of the
solution is gradually varied to modify the optical thickness of the slab. At
low optical thickness, single scattering is dominant and the sample behaves
like a ''diffusive mirror'' : the polarization remains almost unaffected and
the scattered light is detected mainly in the $lin$ $//$ $lin$ channel for
incident linear light, and in the $h\perp h$ channel for circular light. As
optical thickness is increased, higher orders of scattering appear and an
increasing amount of light is redistributed in the orthogonal channels. For
high values of the optical thickness, the light is almost depolarized and the
intensity ratio is close to unity. The fact that the curve for linear
polarization is above that for circular light is probably due to the
contribution of low scattering orders (which is significant even in a
semi-infinite medium\cite{vanDerMark}). Indeed, it is known that the
''memory'' of the initial polarization is preserved longer for linear than for
circular polarization in the case of Rayleigh scatterers \cite{polar}.

\subsection{Enhancement factor}

The accurate determination of the enhancement factor in CBS experiments is
quite delicate \cite{wiersma}. Indeed, the observed enhancement factor is
usually quite smaller than the theoretical prediction of 2. This reduction may
arise from many causes. We have already mentioned the convolution due to the
experimental resolution and the divergence of the probe beam. We also saw, in
the previous section, that the theoretical enhancement factor is smaller than
2 in the $lin\perp lin$ and $h\perp h$ channels. Reciprocity predicts an
enhancement factor of 2 in both the $lin$ $//$ $lin$ and $h$ $//$ $h$
channels. This is assuming that single scattering is eliminated, which is
possible only in the $h$ $//$ $h$ channel (for spherical or Rayleigh
scatterers). Thus, in channels other than $h$ $//$ $h$, the enhancement factor
depends a priori on the sample geometry and optical thickness.

\begin{figure}[h]
\center
\includegraphics[width=\textwidth]{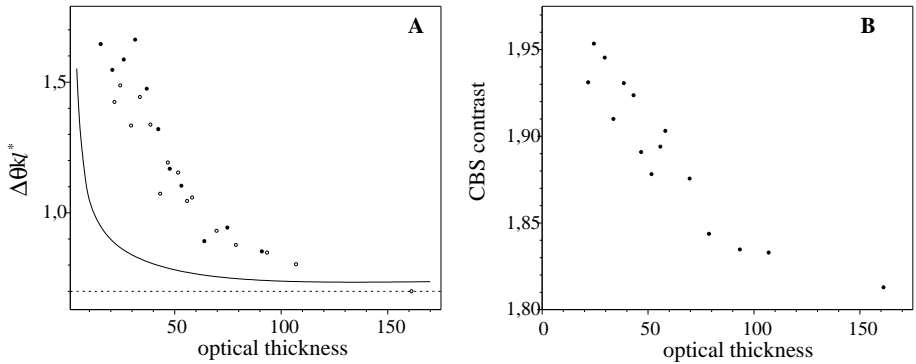}
\caption{Effect of intensity profile on the enhancement factor.}
\label{fig7}
\end{figure}

However, even in the $h$ $//$ $h$ channel, another effect can reduce the
enhancement factor. As emphasized by the Young slits analogy, CBS is
essentially a two-waves interference effect. What determines the contrast of
the interference is the \textit{correlation} between the fields at the input
and output scatterer positions. This correlation includes both differences in
amplitude and phase of the waves at the two points. For instance, the
intensity distribution can be homogeneous and the phase\ vary in the
transverse plane :\ in this situation of partial spatial coherence, the
enhancement factor is decreased \cite{Lenke}\cite{Coherence}.\ In the case of
a gaussian laser beam, the spatial coherence is high and it is rather the
inhomogeneous intensity\ profile that plays a dominant role, as shown
on  fig~\ref{fig7}. If the distance between the first and last scattering event of a given path
is larger than the transverse size of the laser beam $w$, then the amplitudes
of the direct and reverse path are imbalanced and the enhancement factor will
be reduced. One expect the reduction effect to be more important for
increasing values of $l^{\ast}/w$. In most samples $l^{\ast}\ll w$ and this
effect remains small. However, we will see in section $4.3.1$ that in the case
of the atomic sample the above condition is not necessarily fulfilled, and
this reduction effect should be considered.

\subsection{Role of sample geometry}

In the case of a semi-infinite medium, the CBS cone width gives direct access
to the transport mean free path $l^{\ast}$ trough eq.(\ref{FWHMcbs}). However,
in the case of a finite medium, this simple relationship does not hold
anymore, due to the truncation of long light paths. This yields a higher
relative contribution of low scattering orders and hence a broader cone. How
strong this broadening is depends on the actual geometry of the sample. For
instance, in a spherical sample of diameter $\phi$, high scattering orders
will be truncated faster than in a slab of thickness $e=\phi$.

\begin{figure}[h]
\center
\includegraphics[width=\textwidth]{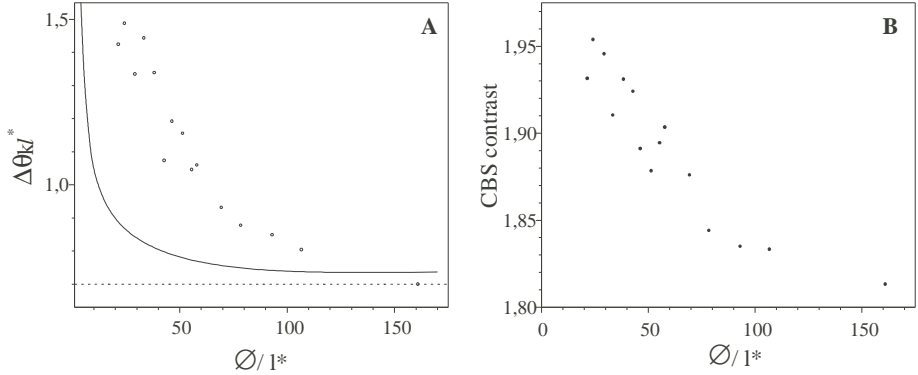}
\caption{CBS experiments on spherical polystyrene samples.
\textbf{A} : measured cone width as a function of $\phi/l^{\ast}$ (circles),
and theoretical cone width for a slab of thickness $\phi$ (line) \textbf{B} :
measured CBS enhancement factor as a function of $\phi/l^{\ast}$.}
\label{fig8}
\end{figure}

To illustrate the importance of sample geometry, we have reported on fig~\ref{fig8} the
results from CBS experiments on spherical samples of polystyrene foam with
different diameters.

In fig~\ref{fig8}\textbf{A} we plotted the product $\Delta\theta_{CBS}kl^{\ast}$ (where
$\Delta\theta_{CBS}$ is the cone's angular FWHM) as a function of the
''optical diameter'' defined as $\phi/l^{\ast}$, where $\phi$ is the diameter
of the sample. The value of $l^{\ast}\approx0.18$ mm was deduced from the
width of cones from bulk samples using eq.(\ref{FWHMcbs}). The circles
correspond to the experiment. The solid line is the prediction of a rigorous
theory \cite{vanDerMark} for scalar waves and a slab geometry ; the horizontal
axis thus correspond , for this curve, to the optical thickness $b=e/l$ where
$e$ is the slab thickness (we assume $l=l^{\ast}$). One can see that the CBS
cone from a spherical sample starts to broaden even at large $\phi/l^{\ast}$
ratio, which reflects the fact that long light paths are truncated faster than
in the slab geometry. We will see that in the case of the atomic sample, the
symmetry is spherical but with a non uniform (quasi-gaussian) density profile;
we thus can expect truncation effects to play an important role in this
situation. In \textbf{B} is reported the measured enhancement factor, which
increases significantly as the sphere's diameter decreases (the peak's height
increases by $\sim15\%$). Two effects tend to increase the enhancement factor.
Due to the truncation of long scattering paths, the cone is broadening and the
convolution by the transfer function of the apparatus is decreased. However,
this does not seem enough to fully explain the observed increase in
enhancement factor. We think that part of this improvement is due to an
increasingly uniform illumination of the sample, reducing the imbalance effect
of fig~\ref{fig7}.

\subsection{CBS with ''single scattering''}

\begin{figure}[h]
\center
\includegraphics[width=\textwidth]{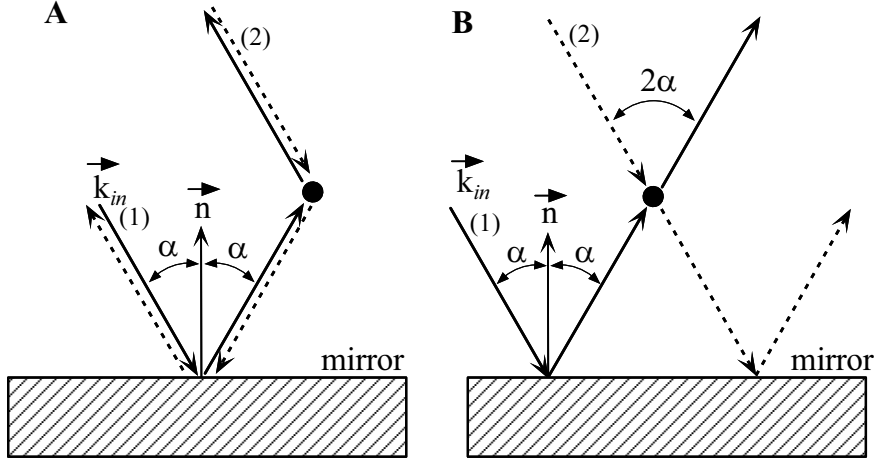}
\caption{CBS with ``single scattering'' : principle.}
\label{fig9}
\end{figure}

We mentioned in section $3.2$\ that single scattering does not contribute to
the CBS signal.\ However, there is a situation where average-robust
interference effects can be observed with single scattering : the case of an
optically thin sample in front of a mirror. This situation is depicted on
fig.~\ref{fig9}.

The scattering medium being optically thin, a certain amount of light reaches
the mirror and is reflected. The mirror plays the role of a second scatterer
with a very anisotropic radiation pattern due to specular reflection.
Fig.~\ref{fig9}\textbf{A} and \textbf{B} illustrate two processes that yield a
constructive interference after configuration average. Process \textbf{A}
corresponds to the ''usual'' backscattering situation, while the example
\textbf{B} shows that the interference is also constructive at an angle
$2\alpha$ from the incident direction. In the far field, this gives a ''ring''
of angular diameter $2\alpha$ for the enhanced scattered intensity, centered
on the direction of the normal to the mirror. The effect can also be
understood as double scattering by the ''real'' scatterer and its image in the mirror.

\begin{figure}[h]
\center
\includegraphics[width=\textwidth]{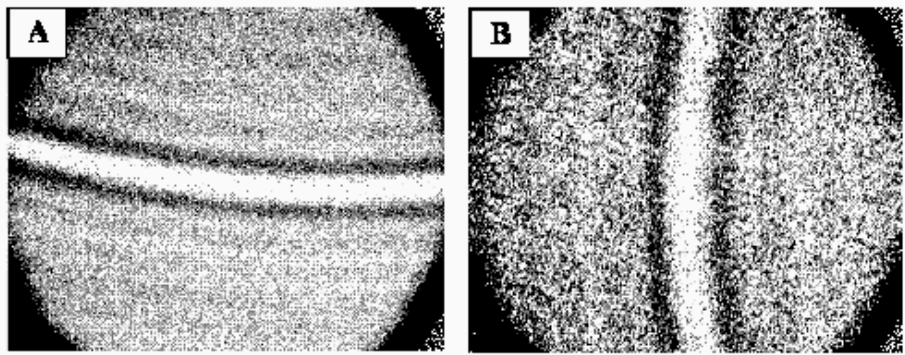}
\caption{CBS with ''single scattering'' : experiment. \textbf{A :
}Mirror tilted vertically \textbf{B} : mirror tilted horizontally.}
\label{fig10}
\end{figure}

This ''single-scattering cone'' can be observed when one performs CBS
experiments on dilute liquid samples in a glass cell of slab geometry. The 4\%
reflection from the back of the cell is enough to yield an important contrast
of the interference, as illustrated on fig.~\ref{fig10} \textbf{A} and
\textbf{B}.

In this experiment, a dilute solution of TiO$_{2}$ particles was placed in a
quadrangular glass cell of thickness 7 mm. The optical thickness of the slab
was $b=0.6$.\ This is not a pure single scattering situation, but higher
scattering orders are not dominant. The images were recorded in the $lin$ $//
$ $lin$ channel, with an incident polarization vertical in the plane of the
figure. In fig.~\ref{fig10} \textbf{A} the cell was tilted vertically so that the vector
normal to the back face points upwards. The image shows the section of the
enhanced intensity ring in a narrow angular range around the backscattering
direction, which is almost an horizontal line. When the cell is tilted
horizontally, we obtain the vertical line of fig.~\ref{fig10} \textbf{B}. To confirm that
the effect arises from scattering, we replaced the solution in the cell by
pure water, and the ring disappeared. Since the process involves only single
scattering (and the reflection from a mirror), the polarization is preserved
around the backscattering direction. Thus, when we recorded the backscattered
intensity in the $lin\perp lin$ channel, the ring also disappeared.

It is thus possible with this configuration to study an interference effect
very similar to CBS, but in the single scattering regime.\ This is an
interesting possibility in the case of an atomic sample, as the theory becomes
much simpler.

\section{Experiments with cold atoms}

\subsection{Properties of atomic scatterers}

Coherent backscattering constitutes a new tool to probe the properties of cold
atoms. Indeed, atoms as elementary scatterers are an interesting medium to
study the quantum manifestations of the interaction between light and matter.
As a consequence of the discrete energy levels, the atom's scattering
cross-section is highly resonant (Q $\approx10^{8}$) and the resonance
frequency is identical for all the scatterers in the sample (assuming a
negligible Doppler effect, which implies laser cooling). Such a situation
would be very difficult to achieve with classical resonators like, for
instance, dielectric spheres of high finesse. Due to this narrow resonance,
the light mean free path in the atomic medium can be varied by orders of
magnitude by shifting the wave frequency a few linewidths away from the atomic
transition. As we will see in section $4.3.1$, the presence of an internal
structure in the ground state (Zeeman sublevels) has some other profound
consequences on the CBS signal from the atomic sample used in our experiment.

Several reasons motivate the use of \textit{cold} atoms to observe CBS.
Firstly, Doppler broadening is then reduced and all the atoms have the same
resonant scattering cross-section, characterized by the natural width $\Gamma$
of the atomic transition ($\Gamma/2\pi\approx$ 6 MHz for rubidium). However,
the atom's motion has a more important consequence on CBS : if the motion of
the scatterers is fast compared to the time for the scattered wave to pass
through the medium, the two reverse waves of fig~\ref{fig1} will encounter different
configurations, resulting in a ''dynamic'' break down of reciprocity. For non
resonant scatterers, the typical time scale to consider is the propagation
time between two scattering events, so this effect requires extremely fast
(almost relativistic) motion \cite{golub}. In the case of quasi-resonant
scattering by atoms, however, the time scale is considerably increased by the
large delay time $\tau=\partial\varphi/\partial\omega$ \cite{delayT}, where
$\varphi$ is the scattered wave phase-shift and $\omega$ the angular
frequency. Taking as a criterion for the break down of the coherent
backscattering cone that each scatterer has moved by one wavelength during
that typical time scale :%

\begin{equation}
\Delta x=v\tau\gtrsim\lambda
\end{equation}
and taking an on-resonant scattering dwell time $\tau_{res}=2/\Gamma$, one
requires to observe CBS velocities smaller than :
\begin{equation}
v_{crit}\backsim\frac{\lambda}{\tau_{res}}=\frac{\lambda\Gamma}{2}%
\label{vitcritique}%
\end{equation}
In terms of Doppler broadening, this corresponds to :%

\begin{equation}
kv_{crit}\backsim\Gamma\label{vcrit}%
\end{equation}

The above criterion shows that if one employs resonant laser light on a dilute
atomic gas (to maximize the optical thickness of the sample and favor multiple
scattering), one has first to laser-cool these atoms in order to observe CBS.
For rubidium atoms, satisfying condition (\ref{vcrit}) implies cooling down
the atomic sample below $T_{crit}=0.25K$, a regime easily reached by standard
techniques. However, in the case of higher orders of scattering, the dwell
time has to be multiplied by the number of scattering events and the above
criterion imposes lower temperatures.

One could however consider the possibility of using atomic gases at room
temperature. But, in order to fulfill (\ref{vitcritique}), one would have to
detune the laser frequency from resonance in order to lower the dwell time.
This would yield the condition :%

\begin{equation}
\tau=\frac{2}{\Gamma}\frac{\left(  \Gamma^{2}\diagup4\right)  }{\delta
^{2}+\left(  \Gamma^{2}\diagup4\right)  }\lesssim\frac{\lambda}{v}%
\end{equation}

or%

\begin{equation}
\left(  \frac{\delta}{\Gamma}\right)  ^{2}\gtrsim\frac{kv}{\Gamma
}\label{deltavit}%
\end{equation}

On the other hand, increasing the detuning will decrease the scattering
cross-section and hence the optical thickness $b\left(  \delta\right)
=n\sigma\left(  \delta\right)  L$ of the sample (where $n$ is the atomic
density, $\sigma$ the scattering cross-section and $L$ the thickness of the
sample). To obtain an optical thickness larger than unity, one has to fulfill :%

\begin{equation}
b\left(  \delta\right)  =b\left(  0\right)  \frac{\left(  \Gamma^{2}%
\diagup4\right)  }{\delta^{2}+\left(  \Gamma^{2}\diagup4\right)  }\gtrsim1
\end{equation}

or%

\begin{equation}
\left(  \frac{\delta}{\Gamma}\right)  ^{2}\lesssim b\left(  0\right)
\label{deltaepaiss}%
\end{equation}

The two conditions \ref{deltavit} and \ref{deltaepaiss} can be fulfilled
simultaneously if :%

\begin{equation}
b\left(  0\right)  \gtrsim\frac{kv}{\Gamma}%
\end{equation}

i.e. if the on-resonance optical thickness $b\left(  0\right)  $ of the medium
is larger than the Doppler broadening in units of $\Gamma$. It seems to be
possible to be realize such situations in hot atomic vapors, but up to now no
coherent backscattering of light by hot atoms has been reported.

\subsection{Preparation of the atomic sample}

The first step in our experiment is to prepare an atomic sample dense enough
to reach the multiple scattering regime. The relevant parameter is the optical
thickness of the atomic cloud $b$. To study multiple scattering of light in
the atomic medium, one\ needs typically $b>1$.

\begin{figure}[h]
\center
\includegraphics[width=0.75\textwidth]{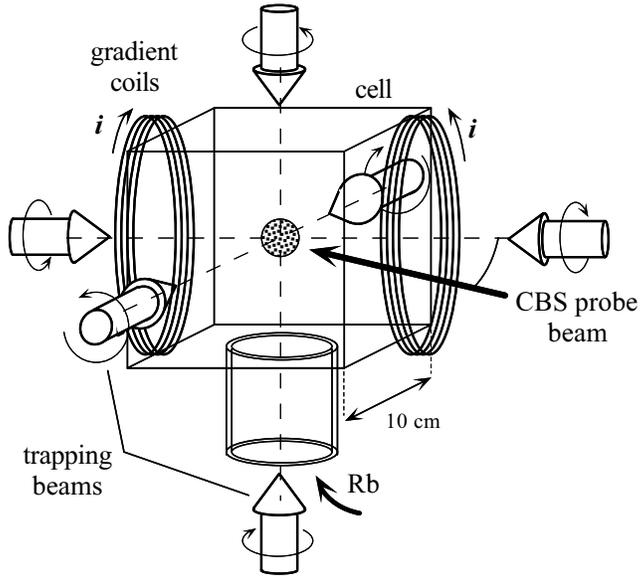}
\caption{Setup of the magneto-optical trap.}
\label{fig11}
\end{figure}

A magneto-optical trap (MOT) is loaded from a room-temperature vapor of
rubidium atoms in a quartz cell, as shown on fig.~\ref{fig11}.

The atoms are trapped by six independent laser beams : this configuration
(instead of the usual three retro-reflected beams) allows to avoid the
imbalance in trapping beams intensity due to the high optical thickness, and
thus to obtain stable trapping. The beams parameters are : wavelength
$\lambda=780$ nm (D2 line of rubidium), detuning from resonance $\delta
\approx-3\Gamma$, diameter 2.8 cm (FWHM) and power per beam 30 mW. The large
beam size increases the number of trapped atoms, but requires more laser
power. These trapping beams are obtained by splitting a single 200 mW beam,
produced\ by single-pass amplification of a 4 mW beam through a tapered
amplifier (SDL TC30-E). The source laser diode is injection-locked to a
reference DBR laser diode (Yokogawa YL78XNW/S). A magnetic field gradient of
typically 10 G/cm is applied to spatially confine the cold atoms. The CBS
laser probe lies in the horizontal plane of the figure, at an angle of
approximately 25$%
{{}^\circ}%
$ from the trapping beam.

To characterize the trap, we record the fluorescence on a photodiode and a CCD
; we also measure the transmission of a the CBS laser probe through the atomic
cloud to determine its optical thickness. The data from these two methods are
in good agreement. The temperature of the cloud is measured by time-of-flight.
The MOT contains approximately $10^{9}$atoms with a quasi-gaussian spatial
distribution of width typically 7 mm FWHM. The MOT loading time is typically
of $0.6\sec$. Transmission measurements using the $3\rightarrow4$ transition
of the D2 line yield a typical optical thickness b $\simeq$ 3. The rms
velocity of the atoms is 10 cm/s, small enough to fulfill the criterion
discussed above.

\begin{figure}[h]
\center
\includegraphics[width=0.75\textwidth]{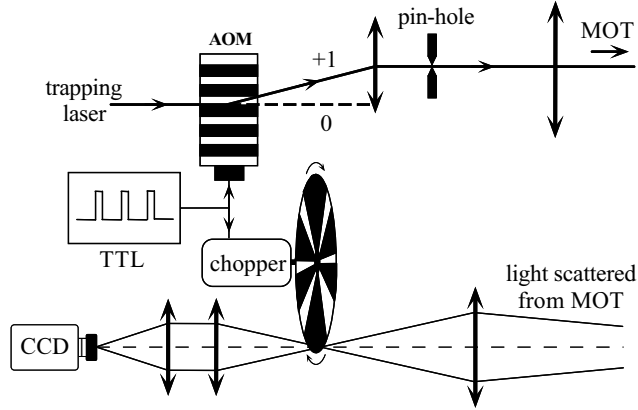}
\caption{Synchronization of the detection.}
\label{fig12}
\end{figure}

To observe the CBS cone, we have to turn off the MOT trapping beams. This is
because the fluorescence of the atoms excited by the trapping lasers (total
scattered power $\simeq$ 4 mW!) is much brighter than the light scattered from
the probe. Also, it seems preferable to avoid perturbations of the atoms by
the trapping lasers during the CBS measurement. Thus, we alternate a ''MOT
phase'' (duration 20 ms) where the atoms are trapped, with a ''CBS'' phase
(2-3 ms) where the trapping beams, repumper and magnetic field are switched
off (switch-off time
$<$%
0,2 ms) and the CBS signal recorded ; this phase is sufficiently short so that
all the atoms remain in the capture zone and are efficiently recaptured when
the MOT is switched back on. In fact, what limits the duration of the CBS
phase is the maximum number of photons that can be scattered by each atom
before it is ''pushed'' out of resonance due to momentum transfer or pumped to
the F $=2$ hyperfine level. For our rubidium atoms, this requires around 1000
photons, which are scattered within 5ms for a saturation parameter
$s_{0}=0.01$. With the ''duty cycle'' described above, the number of atoms in
the trap is stationary and we can chain many such cycles. One problem is that
the CCD camera which detects CBS can not be triggered at such a high rate. The
CCD remains all the time in the ''acquisition'' mode and thus has to be
protected from the bright light scattered during the\ MOT phase. This is
achieved using a chopper wheel as shown on fig.~\ref{fig12}.

The chopper is placed in the focal plane of the analysis lens. A transport
system images this focal plane on the CCD. The trapping laser is turn on an
off with an acousto-optic modulator (residual power 0.2 $\mu$W per beam). The
same TTL signal is used to drive the modulator and as a reference for the
controller of the chopper. The phase is adjusted so that the chopper blades
block the detection path when the trapping laser is on. With this system, we
are able to take exposures up to several tens of minutes. A typical total
exposure time is 1 minute, with a detected flux of photons between 140 and
1400 photons/pixel/s.

It is necessary to acquire a ''background'' image without cold atoms to
subtract stray light. However, this procedure is more critical than in the
case of classical samples, because this stray light originates from different
sources. For instance, one could take the background exposure with no magnetic
field gradient applied during the MOT phase, which prevents the trapping.
However, in this case, a molasse is still operating during the MOT phase that
produces a sample of cold atoms (with density increase in velocity-space). To
avoid this, one need to turn off either the repumper or the trapping beams to
take the background exposure. In this situation, the background signal
originates essentially from scattering of the probe beam by hot atoms
in the cell.
 
\subsection{Results}

\subsubsection{\bigskip Discussion of the atomic CBS signal}

\begin{figure}[h]
\center
\includegraphics[width=0.9\textwidth]{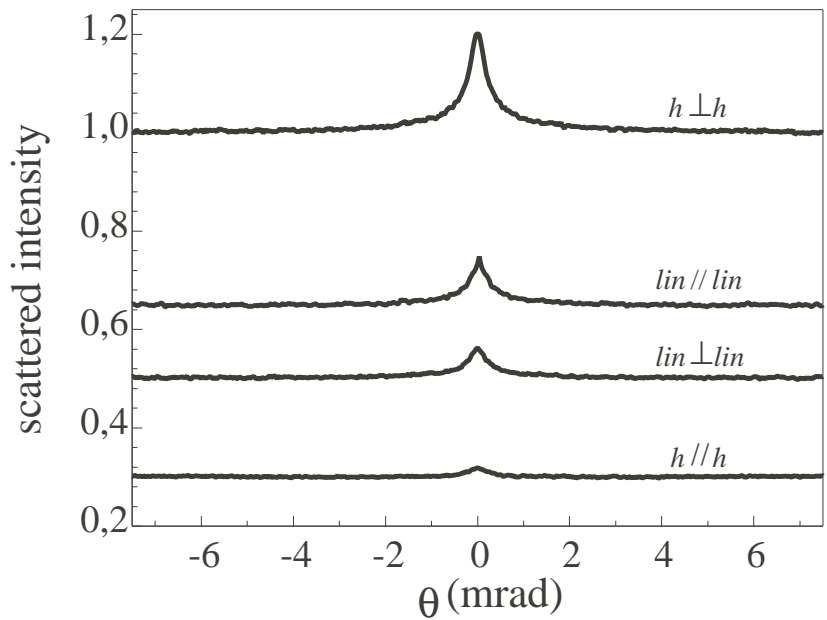}
\caption{Atomic CBS profiles in the four polarization channels.}
\label{fig13}
\end{figure}

Fig.~\ref{fig13} 
shows the profiles of the atomic CBS cones in the four polarization
channels (after angular average).

The detected intensity has been scaled so that the incoherent background of
the $h\perp h$ curve is equal to unity. In this experiment, the on-resonance
optical thickness measured through the center of the trap is $b=3$ for a
quasi-gaussian cloud profile of diameter $\phi\simeq4.8$ mm FWHM, yielding a
peak density of $4\times10^{9}$ cm$^{-3}$ and a scattering mean-free path
$l\simeq1.7$ mm at the center of the trap. A low-intensity probe beam was
used, yielding a saturation parameter $s_{0}=0.01$. The total exposure
(including the ''dark'' periods) lasted 160 s.

\begin{figure}[h]
\center
\includegraphics[width=\textwidth]{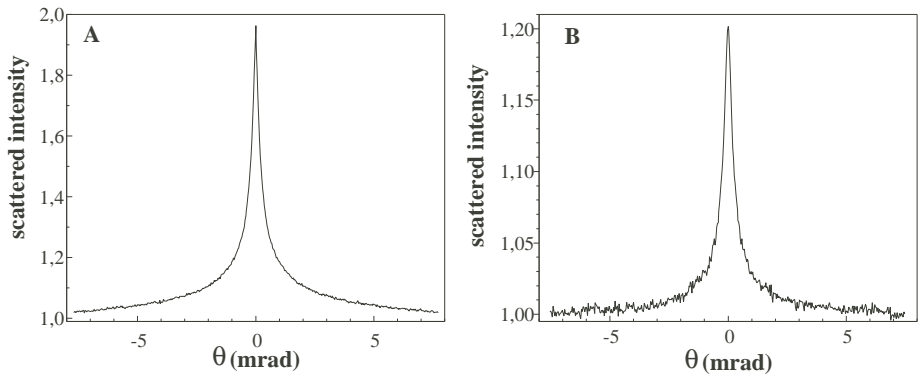}
\caption{CBS cones from \textbf{A} : a sphere of polystyrene foam
($h$ $//$ $h$) \textbf{B} : the atomic cloud ($h\perp h$).}
\label{fig14}
\end{figure}

The values of the enhancement factor are $1.20$ ($h\perp h$), $1.06$ ($h$ $//
$ $h$), $1.15$ ($lin$ $//$ $lin$) and $1.12$ ($lin\perp lin$) respectively.
The cone width, roughly independent of the polarization, is about $0.5$ mrad
FWHM. We thus observe that \textit{the enhancement factor is much smaller than
2} in all polarization channels. Even more striking, the enhancement is only
1.06 in the $h$ $//$ $h$ channel, where reciprocity predicts a value of 2 for
classical (and spherical) scatterers. It is clear that this reduction can not
be attributed to the angular resolution of the apparatus. For a cone width of
0.5 mrad, we expect a reduction of the enhancement by 5\% at most. This is
confirmed by the observation of a CBS cone from a sphere of polystyrene, which
has about the same width as the atomic cone (fig.~\ref{fig14}). 
The enhancement factor
is here of 1.96.

The explanation of this phenomenon lies in the atom's internal structure of
the ground state. First, because our atom is not a two-level system (or a
$0\rightarrow1$ transition), it has a non-negligible probability to make
''spontaneous Raman'' transitions. In such a scattering event, the atom's
internal state (here the Zeeman sublevel of the ground state) after the
scattering is different from the initial state, and the polarization of the
scattered light differs from the incident one. Thus, for most transitions, the
single scattering contribution can not be rejected even in the $h$ $//$ $h$
channel. In this respect, atoms behave similarly to strongly non-spherical
classical scatterers (like e.g. oblate dielectric spheroids).

\begin{figure}[h]
\center
\includegraphics[width=\textwidth]{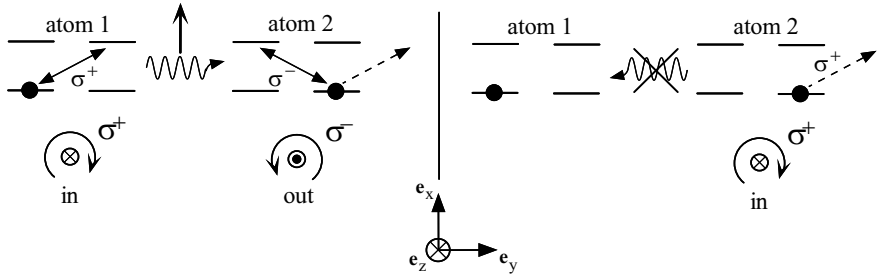}
\caption{Imbalance effect due to the atom's internal structure.}
\label{fig15}
\end{figure}

A more subtle effect is an imbalance between the amplitudes of the reverse
paths that interfere to give the CBS cone. This is illustrated by the simple
example on fig.~\ref{fig15} : we consider double scattering by atoms with a
$1/2\rightarrow1/2$ transition, in the $h$ $//$ $h$ channel. The
quantification axis is taken along the wave vector of the incident laser light
(parallel to $\mathbf{e}_{\text{z}}$). The incident light is polarized e.g.
$\sigma^{+}$ and only the $\sigma^{-}$ component of the scattered light is
detected in the $h$ $//$ $h$ channel.

We suppose that the two atoms are in different Zeeman sublevels. We only
consider the case of Rayleigh scattering (no change of Zeeman sublevel). In
the first path (sequence $1\rightarrow2$, left), atom 1 makes a $\sigma^{+} $
transition and radiates some light toward atom 2 with a linear polarization
parallel to $\mathbf{e}_{\text{x}}$. This is seen by atom 2 as a superposition
of $\sigma^{+}$ and $\sigma^{-}$ light ; since a $\sigma^{+}$ transition is
not available, only $\sigma^{-}$ light is backscattered and detected in the
$h$ $//$ $h$ channel. In the reverse path (right), there is no possibility for
atom 2 to make a $\sigma^{+}$ transition : the amplitude of this path is zero.
This is of course an extreme situation, but calculations\cite{Thibaut}
indicate that this effect reduces strongly the enhancement in the various
channels. These calculations assume double-scattering in a semi-infinite
medium, low saturation, a uniform distribution in the ground state Zeeman
sublevels, and do not include optical pumping effects. Both Rayleigh and Raman
transitions are included, the later also contributing to CBS. A detailed
presentation of these calculations will be reported elsewhere\cite{Thibaut}.
The results for a $3\rightarrow4$ transition are summarized below :%

\[%
\begin{tabular}
[c]{|c|c|c|c|c|}\hline
& $h$ $//$ $h$ & $h\perp h$ & $lin$ $//$ $lin$ & $lin\perp lin$\\\hline
$\gamma_{c}$ & $0.028$ & $0.154$ & $0.108$ & $0.075$\\\hline
$\gamma_{l}$ & $0.131$ & $0.216$ & $0.180$ & $0.167$\\\hline
$\gamma_{s}$ & $0.040$ & $0.510$ & $0.348$ & $0.201$\\\hline
$enhancement$ & $1.166$ & $1.213$ & $1.204$ & $1.206$\\\hline
\end{tabular}
\]
This table contains the contributions to the bistatic coefficient
\cite{bistatic} of the ''crossed'' (or interference) term $\gamma_{c}$,
''ladder'' (or incoherent) term $\gamma_{l}$, and single scattering term
$\gamma_{s}$ at exact backscattering ($\theta=0$). The effective enhancement
factor in the presence of single scattering is then :
\begin{equation}
enhancement=\frac{\gamma_{c}+\gamma_{l}+\gamma_{s}}{\gamma_{l}+\gamma_{s}%
}=1+\frac{\gamma_{c}}{\gamma_{l}+\gamma_{s}}\label{enhancement}%
\end{equation}

Even though the model considers only double scattering and a semi-infinite
medium, the values of \ the table above are surprisingly close to the
experimental observations. They reproduce the order of magnitude of the
enhancement factor and even the hierarchy between the different channels (for
instance, the enhancement is predicted to be smallest in the $h$ $//$ $h $
channel). Note that the reduction of the CBS enhancement factor has different
origins in different channels : in the $h$ $//$ $h$ channel, most of the
reduction stems from the imbalance mechanism of fig~\ref{fig15} ($1+\gamma_{c}%
/\gamma_{l}=1.214$), while a strong single scattering contribution explains
most of the enhancement reduction in the\ $h\perp h$ channel.

The angular width of the atomic cone is $\Delta\theta_{CBS}\approx0.5$ mrad.
This value is about 10 times larger than what is obtained with
eq.(\ref{FWHMcbs}) and the estimated mean-free path of $1.7$ mm at the center
of the trap. This is not surprising, since our sample is very far from a
semi-infinite medium. Because of the spherical symmetry, gaussian density and
rather modest optical thickness of the cloud, low orders of scattering are
expected to dominate,\ yielding a broader cone. A Monte Carlo simulation is
being developed to quantitatively address the problem of our particular sample geometry.

\subsubsection{ Effect of cloud density}

\begin{figure}[h]
\center
\includegraphics[width=\textwidth]{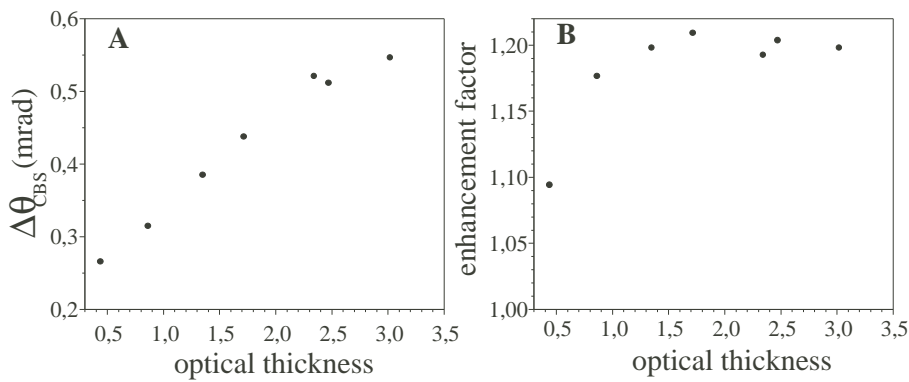}
\caption{Effect of atomic cloud density on the CBS signal ($h\perp
h$). \textbf{A} : cone width as a function of the optical thickness at the
center of the cloud \textbf{B} : enhancement factor.}
\label{fig16}
\end{figure}

By varying the trap parameters e.g. the magnetic field gradient (during the
MOT phase), we can modify to a certain extent the characteristics of the
atomic cloud (size and density). This modifies the width of the CBS\ cone, as
shown on fig.~\ref{fig16}\textbf{A}.

In this experiment, the value of the magnetic field gradient applied during
the MOT phase was varied; this acts on the number of trapped atoms and on the
size of the cloud. The optical thickness and fluorescence profile of the trap
where recorded for each value of the gradient. We can see (\textbf{A}) that
the cone broadens consequently as the optical thickness increases, as one
would expect from a decreasing scattering mean-free path. However, a more
detailed analysis (including the contributions of different scattering orders)
is needed to quantitatively understand these data. Plot \textbf{B} of fig~\ref{fig16}
shows the enhancement factor as a function of optical thickness. It remains
constant except for small values of the optical thickness where it decreases
sharply. Several effect probably contribute to this reduction : increased
weight of single scattering, convolution by the angular response of the apparatus.

\subsubsection{Dipole vs. isotropic radiation pattern}

\begin{figure}[h]
\center
\includegraphics[width=\textwidth]{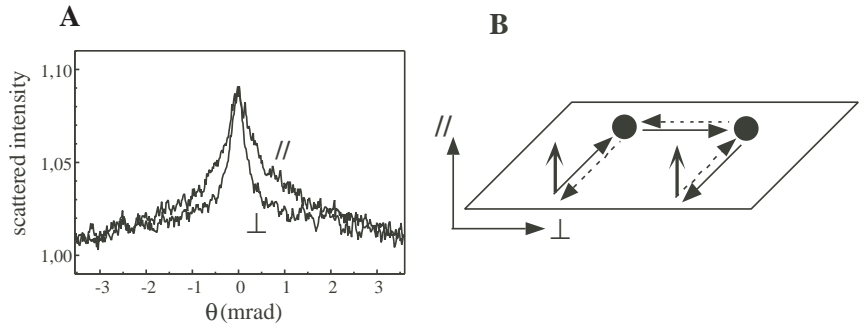}
\caption{Anisotropy\textbf{\ }of the CBS cone in the $lin$ $//$
$lin$ channel. \textbf{A} : cone cross-sections parallel and orthogonal to the
incident polarization \textbf{B} : origin of the anisotropy.}
\label{fig17}
\end{figure}

We have mentioned above that the CBS cone is not always isotropic. Indeed, we
observed some anisotropy on the atomic signal recorded in the $lin$ $//$ $lin$
channel. This is illustrated on fig.~\ref{fig17}\textbf{A} where we reported two
cross-sections of the cone : \textbf{(}$\mathbf{//}$\textbf{)} cross-section
parallel to the direction of the incident polarization, and \textbf{(}%
$\mathbf{\bot}$\textbf{) }cross-section orthogonal to the polarization. The
first profile is clearly wider (by approximately a factor of 2).

This effect, which has already been reported with classical Rayleigh
scatterers\cite{depol}, is due to the combination of low scattering orders and
a dipole-type radiation pattern for the atomic scatterer. Indeed, if we
consider for instance only double scattering and an incident vertical linear
polarization, most of the scattering paths will lie in the horizontal plane
because very little light will be radiated in the vertical direction
(fig~\ref{fig17}\textbf{B}). Thus, the phase difference between reverse paths will vary
much more slowly in the vertical angular direction (where the detector moves
along a fringe of the equivalent Young's interference pattern) than in the
horizontal one (motion orthogonal to the fringe system), yielding an
asymmetric cone. A model including only double scattering (for a semi-infinite
medium) by atoms with an internal structure yields a cone with an asymmetry
close to the experimental observation. Since this effect originates
essentially from the lowest scattering orders, this agreement is not surprising.

\section{Conclusion}

In this paper, we discussed in details our experiment of coherent
backscattering of light from cold atoms. A particular attention was drawn to
the influence of sample and laser probe geometry on the CBS signal, as
illustrated by experiments on classical samples. The small enhancement factors
observed in the experiment on cold atoms are explained by two effects due to
the atom's internal structure : the presence of single scattering (spontaneous
Raman transitions), and a more interesting imbalance effect in the amplitudes
of the time-reversed paths. We are currently setting up a Monte Carlo
simulation to take into account the specific geometry of our sample together
with the internal structure properties of the atomic scatterer. Once this
necessary tool is developed, we plan to quantitatively study the effect of
various parameters such as different atomic transitions, laser frequency and
intensity, or an applied magnetic field.

{\em Acknowledgement\/}:
We thank the CNRS and the PACA\ Region for financial support. We gratefully
acknowledge the important contributions of D.\ Delande and T. Jonckheere to
the theoretical work and numerical simulations, and of J.-C. Bernard to the
development of the experiment.

\end{document}